\documentclass[aps,prb, onecolumn, nofootinbib, citeautoscript,10pt]{revtex4-2}  
\usepackage{dcolumn}
\usepackage{bm}
\usepackage{amsmath}
\usepackage{tabularx, graphicx}
\usepackage{epstopdf}
\usepackage{latexsym}
\usepackage{amssymb}
\usepackage{amsmath}
\usepackage{color, colortbl}
\usepackage{psfrag}
\usepackage{bbm}
\usepackage{titlesec}
\usepackage{dsfont}
\usepackage{slashed}
\usepackage{multirow}
\usepackage[tight]{subfigure}
\usepackage[papersize={8.5in,11in}]{geometry}
\usepackage{graphicx}

\usepackage{color}
\definecolor{darkblue}{rgb}{0.,0.,0.4}
\definecolor{darkred}{rgb}{0.5,0.,0.}
\definecolor{BlueViolet}{RGB}{138,43,226}
\definecolor{SkyBlue}{RGB}{30,144,255}
\definecolor{DarkGreen}{RGB}{0,100,0}
\usepackage[pdftex,colorlinks=true,linkcolor=darkblue,citecolor=blue,urlcolor=darkred]{hyperref}

\usepackage{physics}

\newcommand{\bs}[1]{\boldsymbol{#1}}

\def \nn{\nonumber \\}

\begin{document}

\title{Linear response from tilted Dirac cones under strain-induced pseudomagnetic fields}

\author{Sanskar Sharma}
\author{Ipsita Mandal}
\email{ipsita.mandal@snu.edu.in}

\affiliation{Department of Physics, Shiv Nadar Institution of Eminence (SNIoE), Gautam Buddha Nagar, Uttar Pradesh 201314, India}

\begin{abstract}
We investigate the transport signatures of pseudo-Landau levels (PLLs) in two-dimensional anisotropic Dirac systems with tilted cones, whose effective bandstructure results from strain-induced pseudogauge fields. In contrast to conventional Landau quantisation, the PLLs exhibit explicit momentum-dependence by being dispersive, leading to finite longitudinal group-velocities. We analyse the transport properties within the semiclassical Boltzmann framework by computing the electrical, thermoelectric, and thermal response in the linear regime, which acquire nonzero longitudinal components. We also check the validity of the Mott relation and Wiedemann-Franz law in our system. Our results provide a unified framework for understanding the interplay between tilted spectrum and structural deformation in affecting quantum transport, and suggest unambiguous experimental signatures in strain-engineered systems.
\end{abstract}

\maketitle

\tableofcontents


\section{Introduction}

\begin{figure*}[t]
\centering
\subfigure[]{\includegraphics[width = 0.23 \textwidth]{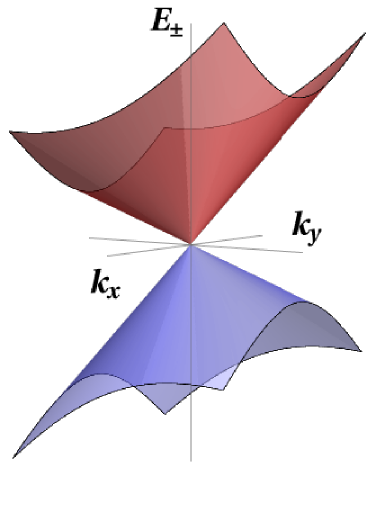}} \hspace{4 cm}
\subfigure[]{\includegraphics[width=0.5\textwidth]{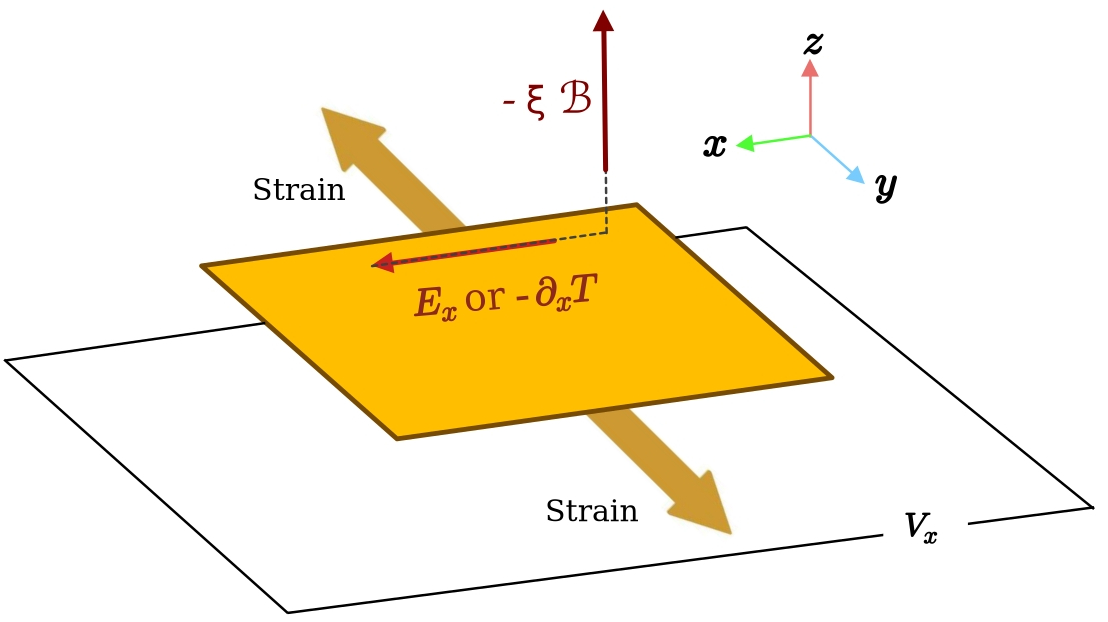}}
\caption{\label{figdis}(a) Energy bands for a tilted Dirac cone. (b) Set-up to measure the longitudinal transport coefficients in the presence of a strain-induced pseudomagnetic field. Here, a uniform and time-independent electric field ($E_x$) or temperature gradient ($- \,\partial_x T$) is applied in the $x$-direction and $V_x$ is the associated voltage.}
\end{figure*}

The exceptionally novel electronic properties of graphene, a two-dimensional (2d) avatar of the allotropes of carbon, have contributed to exotic quantum-transport phenomena, which can be attributed to the emergent Dirac cone dispersions. The Dirac cones appear at the $K$ and $K^\prime $ points of the hexagonal Brillouin zone (BZ), when the Fermi level is set at the charge-neutral point of the system. BEDT-TTF- and BEDT-TSeF-based quasi-2d organic conductors under hydrostatic pressure, such as $\alpha$-(BEDT-TTF)$_2$I$_3$ and $\alpha$-(BEDT-TSeF)$_2$I$_3$ [where BEDT-TTF and BEDT-TSeF stand for bis(ethylenedithio)tetrathiafulvalene and bis(ethylene)dithiotetraselenafulvalene, respectively], also possess Dirac cones like graphene Ref.~\cite{organic-tilted-dirac, organic-dirac}. Usually, the pressure-dependent bandstructure exhibits a large anisotropy, e.g., tilted spectrum Ref.~\cite{organic-tilted-dirac}. Electronic quasiparticles in the vicinity of a single cone can be described by an effective Weyl-like Hamiltonian in 2d. Using an effective tight-binding model on an anisotropic triangular lattice, with two sublattice sites (viz. A and B), one can show that the tilting of the Dirac cones [cf. Fig.~\ref{figdis}(a)] is caused by next-nearest-neighbour (nnn) hopping terms. A quinoid-type lattice-deformation can cause a tilt in the Dirac cones of graphene as well. As argued in Ref.~\cite{tilted-goerbig}, a necessary requirement for the nnn hoppings to cause a tilt in the form of a linear-in-momentum term is that the Dirac-points (which are the locations of the Dirac cones) be situated at points in the first BZ which do not coincide with the high-symmetry points (e.g., the corners $K$ and $K^\prime$ of the graphene lattice). In fact, if the Dirac-points (say, $D$ and $D^\prime $) coincide with the $K$ and $K^\prime$ points, then a quinoid-type graphene will have the nnn terms contributing only at the quadratic-in-momentum corrections.
Another point is that a tilt in the Dirac cones of quinoid-type graphene will be less pronounced than that in the organic conductors, because the strength of the nnn terms is of the same order of magnitude as that of the nearest-neighbor (nn) hopping terms in the latter Ref.~\cite{tilted-goerbig}. The reason behind this is the following: The locations of the Dirac cones naturally deviate from high-symmetry crystallographic points in the organic compounds under pressure, leading to the nnn hopping terms becoming significant. One can capture the anisotropic features of the above systems via a generalized Weyl Hamiltonian that incorporates the tilt-parameters and any other anisotropy Ref.~\cite{tilted-goerbig-plasmons, strained-dirac-franz, ips_tilted_dirac}. In particular, the nnn hoppings also make the Fermi velocities ($v_x $ and $ v_y$) along the mutually-perpendicular momentum-axes to be unequal (i.e., $ v_x \neq v_y$).

Elastic strain has been used to manipulate the electronic properties of such 2d systems hosting Dirac cones, opening up unprecedented opportunities for engineering quantum states through structural modifications Ref.~\cite{neto-strain-graphene, strain-2d, graphene-strain-review, graphene-strain-geim, graphene-strain-terms, graphene-pseudo0, graphene-pseudo}. One of the remarkable outcome is the creation of strain-induced pseudomagnetic fields Ref.~\cite{graphene-pseudo0, graphene-pseudo, strained-dirac-franz, complex-LL-strained-graphene}, which would cause pseudo-Landau levels (PLLs) to appear, analogous to the Landau-levels (LLs) appearing under the action of an actual magnetic field. The PLLs provide a route to realise the physics of Landau-level-quantization even under a zero magnetic field.The coupling of the fermionic quasiparticles with the mechanical strain is described using the pseudogauge field, $\boldsymbol{\mathcal A}$, where the latter is generated by the lattice deformations. The pseudovector potential thus acts upon the charge carriers as if they were subjected to a magnetic field, $\boldsymbol{\mathcal B} = \nabla_{\boldsymbol r} \times \boldsymbol{\mathcal A}$~\cite{pikulin16_chiral, ips_rahul_ph_strain, ips-ruiz, ips-rsw-ph, ips-nlsm-strain}. It is important to note that the nature of the coupling differs from that of an actual gauge field, the former being valley-dependent, where we refer to $D$ and $D^\prime $ as the two inequivalent valleys. The sign of the coupling switches for $D$ and $D^\prime $, as shown in Ref.~\cite{strained-dirac-franz}. The situation is similar to the 3d versions embodied by a pair of conjugate Weyl nodes with opposite chiralities Ref.~\cite{pikulin16_chiral}.

In this paper, we will study the nature of the bulk linear response arising from the dispersive PLLs of tilted Dirac cones. The schematic experimental set-up is shown in Fig.~\ref{figdis}(b). Expanding upon earlier works \cite{strained-dirac-franz, complex-LL-strained-graphene}, our main aim is to quantify the effects arising from the additional ingredient of tilting the spectra.

\section{Model}
\label{secmodel}


The minimal low-energy effective Hamiltonian of a single tilted Dirac cone, at a valley indexed by $\xi $ (where $\xi =\pm$), is captured by Ref.~\cite{tilted-goerbig, ips_tilted_dirac}
\begin{align}
\label{eqham}
\mathcal{H}_0 =   \xi\, v_x\, k_x \,\sigma_x +    v_y \, k_y \,\sigma_y
+ \left ( \boldsymbol{v}_0 \cdot\boldsymbol{k} \right ) {\mathbb I}_{2\times 2} \,, \quad
\boldsymbol{v}_0 \equiv  
v_{0x} \,   {\boldsymbol{\hat x}} +  v_{0y} \,  {\boldsymbol{\hat y}}\,, \quad
v_{0x} \equiv   v_x\,\eta_x \, ,\quad
v_{0y} \equiv  v_y\,\eta_y \,,
\end{align}
where $\sigma_x $ and $\sigma_y $ denote the first two Pauli matrices. The velocities, $v_x$ and $v_y$, represent the components of the Fermi velocities along the $k_x$- and $k_y$-axes, respectively. Finally, the amount of tilt with respect to these two axes
are parametrised by $\eta_x$ and $\eta_y$, respectively. Clearly, the tilt introduces diagonal terms in the $2\times 2$ Hamiltonian, which did not exist in the untilted Dirac cone case.
In the following, we will use the indices $\lbrace i, \, j \rbrace $ to label the component of a vector (or tensor) along the $
\lbrace i^{\rm th}, \, j^{\rm th} \rbrace$ Cartesian coordinate.

Here, we restrict ourselves to the cases when $ \eta <1$, where $\eta  \equiv \sqrt{ {\eta_x}^2 + {\eta_y}^2  }$, which implies that we are going to consider the type-I (or undertilted) phases, where the 1d Fermi surfaces at finite values of the chemical potential ($\mu$) are closed, i.e., ellipses. The energy eigenvalues of the two-band Hamiltonian are:
\begin{align}
E_{\pm} ({\boldsymbol k}) = (\boldsymbol{v}_0 \cdot \boldsymbol{k} )
 \pm \, \epsilon_k \,, \text{ where } 
\epsilon_k = \sqrt{ v^2_x \, k^2_x + v^2_y \, k^2_y } \,.
\end{align}

\subsection{Incorporating strain}


The real-space Hamiltonian in the vicinity of a Dirac cone, containing only nearest-neighbour (nn) hoppings, can be expressed as
\begin{align}
\label{eqnn}
& H_d = -\, t \sum_{\bs r_A} \sum \limits_{ \lbrace \bs M_\alpha \rbrace } \left[ 
a_{\bs r_A}^\dagger \, b_{\bs r_A + \bs M_\alpha} + b_{\bs r_A + \bs M_\alpha }^\dagger \, a_{\bs r}
\right ], \quad \alpha \in \lbrace 1, \,2 ,\, 3 \rbrace \,,\nn
& \bs M_1 = a_0 \, \bs{\hat y}\,, \quad \bs M_2 = - \frac{a_0}{2} \left(\sqrt 3 \,\bs{\hat x} + \bs{\hat y}\right),
\quad \bs M_3 =  \frac{a_0}{2} \left(\sqrt 3 \,\bs{\hat x} - \bs{\hat y}\right),
\end{align}
where $a^\dagger_{\bs r}$ and $b ^\dagger _{\bs r}$ stand for the creation operators associated with the A and B sublattice sites, respectively, which are separated by the nn-distance $a_0$. The discrete coordinate-locations, $\bs r_A$ denote the A-sublattice sites and the set $\lbrace \bs M_\alpha \rbrace$ represents the nn-vectors for hopping between the nn sublattice-sites. We use the effective forms derived in Refs.~\cite{neto-strain-graphene, strained-dirac-franz} for a Dirac cone subjected to uniaxial strain, which produces the pseudomagnetic fields that we want to analyze in this paper. In this procedure, the uniaxial strain pattern is incorporated in the system through a modulation of the nn hoppings into a  smooth function of the spatial coordinates. For the positions $\bs r_A $ and $\bs r_a + \bs M_\alpha $, the hopping amplitude is given the form of 
\begin{align}
t \rightarrow t_{\alpha}(\bs r_A) 
 = t_\alpha\, e^{-\gamma \,\Delta u_\alpha }
=  t  \left ( 1 -  \gamma_a \right)
 + \order{ \gamma _\alpha^2 } \,, \quad \gamma_\alpha \equiv \gamma\, \Delta u_\alpha, 
\end{align}
which depends on a displacement, $ \Delta u_{\alpha}$, along the three nn-bond directions. We assume the original displacement-field is $\bs u (\bs r) = u_x (\bs r)  \, \bs {\hat x} + u_y (\bs r) \, \bs {\hat y}$, such that $\bs u ({\bs r_A} + {}\bs M_\alpha) - \bs u (\bs r_A) = \Delta u_\alpha $, which we postulate to be of small magnitude (such that $\gamma_\alpha \ll 1 $).
Next, we are interested in a configuration of the strain tensor, $\varsigma_{ij}$, such that only its $\varsigma_{yy}$-component survives (i.e., $ \varsigma_{xx} = \varsigma_{xy} = \varsigma_{yx} = 0 $), which is purely $y$-dependent. Invoking the relation $\Delta u_{\alpha} = \sum_{i, j} \frac{ \left(M_\alpha \right)_i \left( M_\alpha\right)_j } {a_0}\,\varsigma_{ij}$, we deduce that
\begin{align}
\Delta u_1 = 4\, \Delta u_2 = 4 \, \Delta u_3 = a_0\, \varsigma_{yy}  (y)\,.
\end{align}

Although the authors in Ref.~\cite{strained-dirac-franz} derived the strain-induced terms for untilted isotropic Dirac cones, initially located at the $K$ and $K^\prime $ corners of graphene's BZ, we assume that this applies also to the case of the tilted versions (shifted away from the high-symmetry points). This is justified because the Dirac-point and the $K$-point of the deformed
lattice do not coincide in general Ref.~\cite{neto-strain-graphene}, which can be attributed to the fact that the applied strain not only affects the hoppings but also causes a distortion of the lattice itself. We just need to ensure that the strain to be applied in such a way that it affects mainly the off-diagonal terms arising from the nn hoppings. Therefore, in the vicinity of a Dirac-point hosting a tilted cone with anisotropic Fermi velocities as well, the applied strain modifies the continuum version of the low-energy Hamiltonian to Refs.~\cite{strained-dirac-franz, complex-LL-strained-graphene}
\begin{align}
\label{eqhtot}
& \mathcal H_\xi (k_x, y) = \left[ 
\left(v_x \, \eta_x \, k_x + v_y \,\eta_y\, k_y \right)  {\mathbb I}_{2\times 2} 
+ \mathcal{H}_s(y; k_x, k_y) \right ]\big \vert_{k_y \rightarrow -i\, \partial_y}
\,, \quad 
\mathcal{H}_s  = v_x \, d^\xi_x (\bs k)\,\sigma_x + v_y\,d^\xi_y (\bs k)\,\sigma_y  \,, \nn &
\nn & d^\xi_x(\boldsymbol{k}) = \xi \left( \left (1 - \frac{ \tilde \gamma} {4}\right ) {k}_x 
+\, \xi \, \frac{ \tilde \gamma} {2\, a_0}\right)  \,,
\quad  d^\xi_y(\boldsymbol{k}) =
\left (1 - \frac{3\, \tilde \gamma } {4} \right )  k_y 
+ \xi\, a_0\,\frac{ k_x \, k_y}{2} \left (1 - \frac{\tilde \gamma} {4}  \right ) ,
\quad \tilde \gamma = \gamma\,a_0 \, \varsigma_{yy}\,.
\end{align}
Here, the index $\xi =\pm$ labels two inequivalent valleys. In this paper, we will consider tilt with respect to the $k_x$-axis only, setting $\eta_y = 0 $.

We observe that the $\varsigma_{yy}$ terms couples to the $k_x$ terms in a way such that the coupling of fermions with the effective pseudogauge-field component, $ \xi\, e \,  \mathcal{A}_x = \xi \, \tilde \gamma /(2\, a_0) = \xi\, \gamma \, \varsigma_{yy} / 2 $ (because, a gauge-field coupling is obtained by the substitution $k_x \rightarrow k_x - \text{charge } \times \mathcal{A}_x $). Thus, $ \mathcal{A}_x $ appears with opposite signs for $ D$ and $D^\prime$. In addition, we find that $\varsigma_{yy}$ also renormalizes the values of the Fermi velocities. For a clearer visualization of the gauge-field coupling, remembering that $  
\tilde \gamma  = 2\, a_0 \,e \,  \mathcal{A}_x $, we rewrite the Hamiltonian with
\begin{align}
 & d^\xi_x(\boldsymbol{k}) = \xi  \left (1 - \frac{ a_0 \,e \,  \mathcal{A}_x} {2}\right ) {k}_x 
+  \xi^2\,  e \,  \mathcal{A}_x  \,,
\quad d^\xi_y(\boldsymbol{k}) =
\left ( 1- 2\,a_0 \,e \,  \mathcal{A}_x  +  \frac{ \xi^2 \,a_0 \,e \,  \mathcal{A}_x} {2} 
\right )  k_y 
+  a_0\,\frac{ \xi\,k_x \, k_y}{2} \left (1 - \frac{ a_0 \,e \,  \mathcal{A}_x} {2}  \right ) \nn
\Rightarrow & \; d^\xi_x(\boldsymbol{k}) = \xi  \,\Pi_x
  - \xi\,k_x  \frac{ a_0 \,e \,  \mathcal{A}_x} {2} \,,
\quad d^\xi_y(\boldsymbol{k}) =
\left ( 1- 2\,a_0 \,e \,  \mathcal{A}_x \right )  k_y  
+  \frac{ a_0\,\xi \, k_y} {2} \,\Pi_x \,,
\text{ where } \Pi_x =  k_x +  \xi \,  e \,  \mathcal{A}_x \,.
\end{align}
We find that a direct interpretation of the Hamiltonian as an analogy of actual LLs is a bit problematic because we have a mix of both $\Pi_x$ and $k_x $ terms. There has been the interpretation of the extra factors as a modulation of the Fermi velocities Ref.~\cite{complex-LL-strained-graphene}, but we will take the point of view of solving for the energy levels of the resulting system and work out its eigenspectra, which we will call as the PLLs.

By taking $\mathcal{A}_x = \, y\, {\mathcal B} $ (analogue of the Landau gauge), we can engineer $\bs{\mathcal B}= \nabla_{\bs r} \cross \bs{\mathcal A }$ to be directed perpendicular to the plane of the 2d lattice, which we call the negative $z$-direction. Therefore, we define
\begin{align}
 \varsigma_{yy} = \frac{2 \, b\, y} {\gamma} \,, \quad
b = e\, \mathcal B\,,
\end{align}
we generate $ \bs{\mathcal B} = -\,{\mathcal B} \, \bs{\hat z}$, which acts on the two valleys with opposite signs. Because of breaking the translation symmetry along the $y$-direction, $k_y $ is no longer a good quantum number, and we are forced to work in a hybrid spatial-momentum coordinates (viz. $y$ and $k_x$). Overall, our system acquires anisotropy from three ingredients: (i) $ v_x \neq v_y$; (ii) uniaxial strain; (iii) tilting. 

\begin{figure*}[t!]
\centering
\subfigure[]{\includegraphics[width= 0.7 \textwidth]{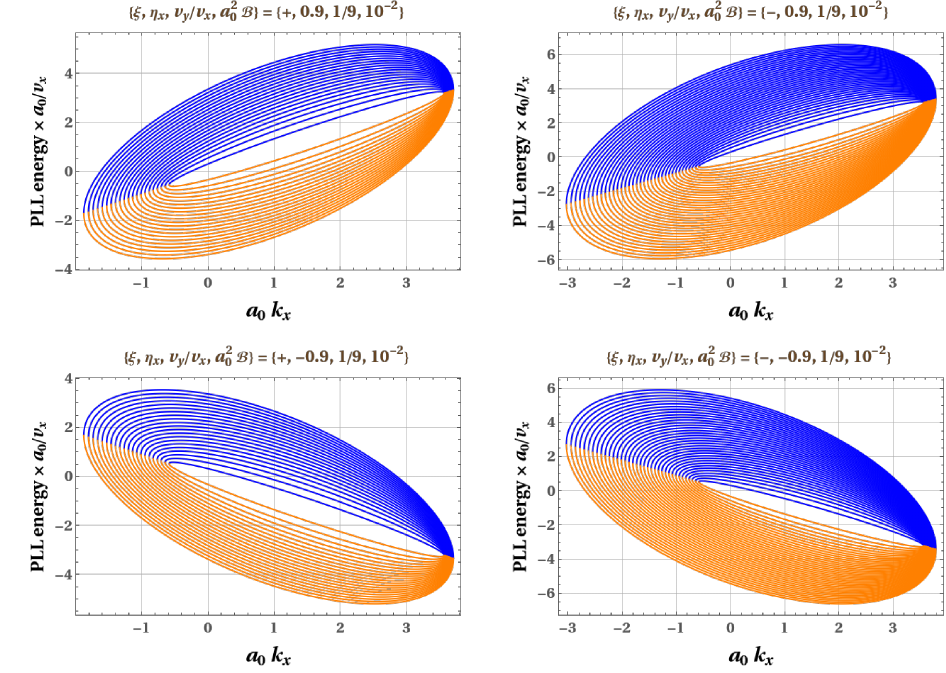}}
\subfigure[]{\includegraphics[width= 0.7 \textwidth]{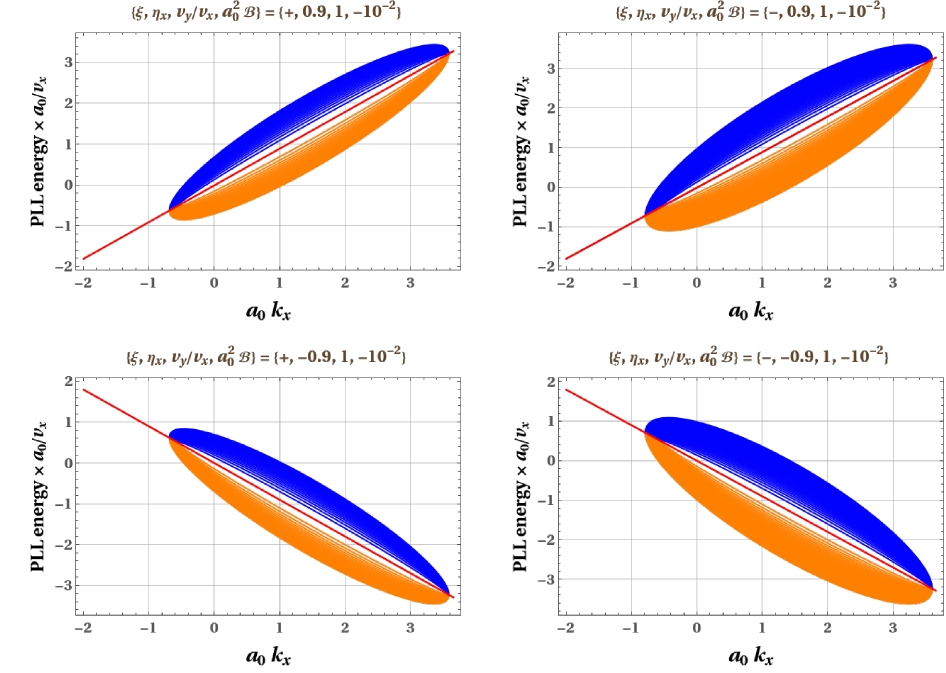}}
\caption{Dispersion of the PLLs for some chosen parameters shown in the plot-labels. \label{fig:spectrum_B}}
\end{figure*}

\subsection{Pseudo-Landau levels}
\label{secll}

\begin{figure*}[th!]
\centering
\includegraphics[width= 0.7 \textwidth]{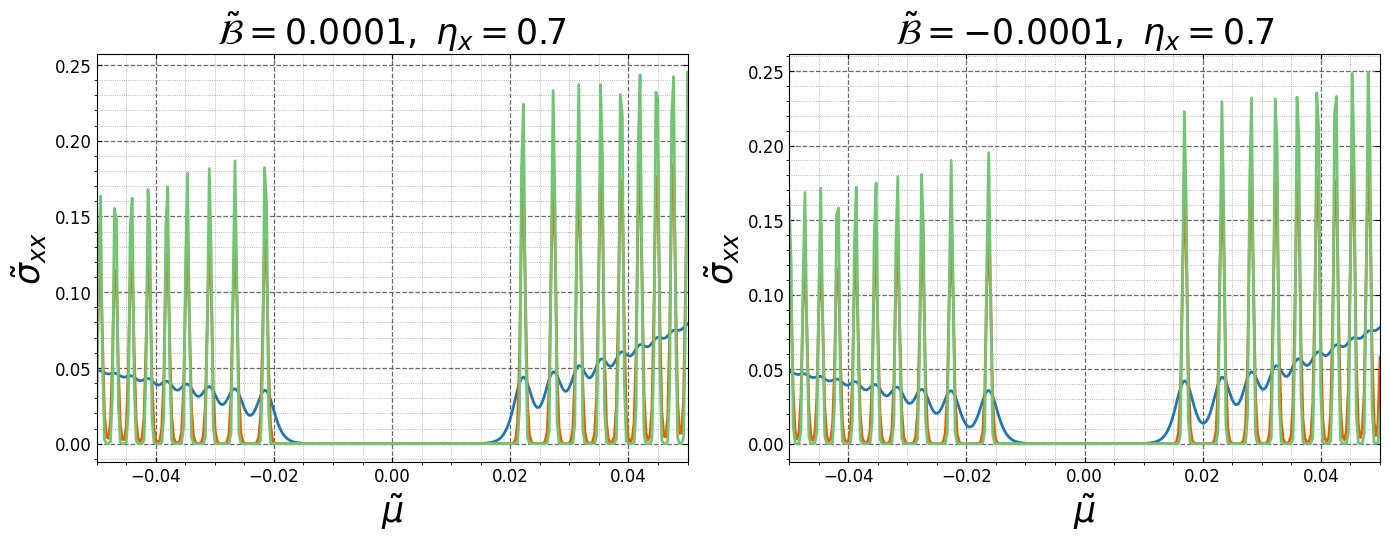}
\includegraphics[width=0.7\textwidth]{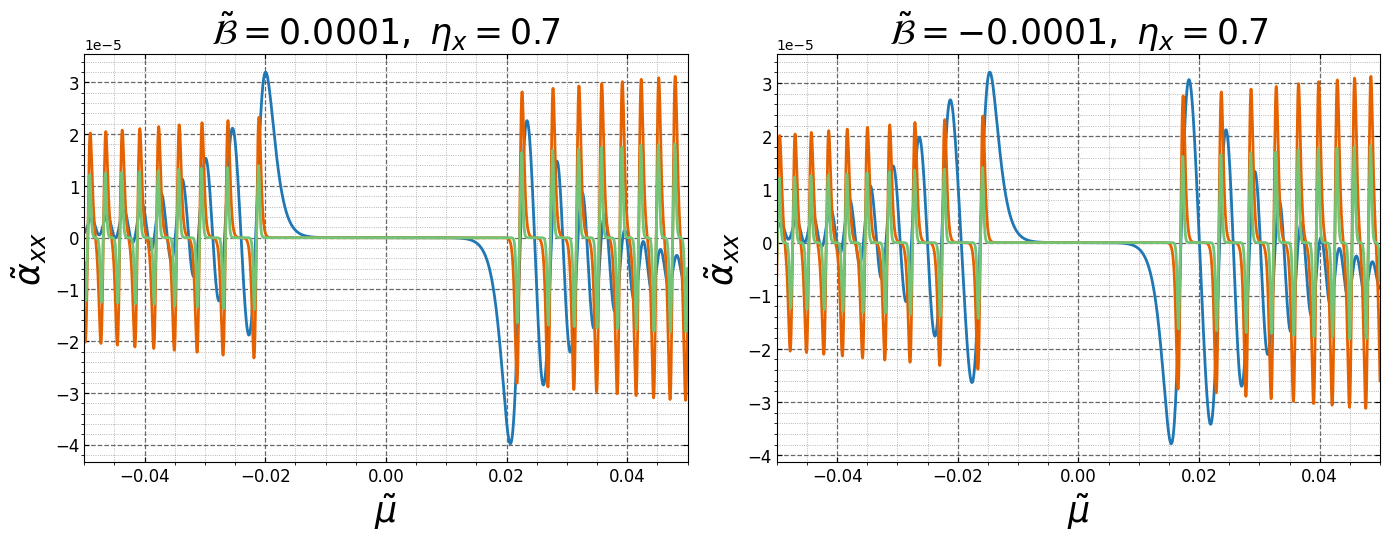} 
\includegraphics[width=0.7\textwidth]{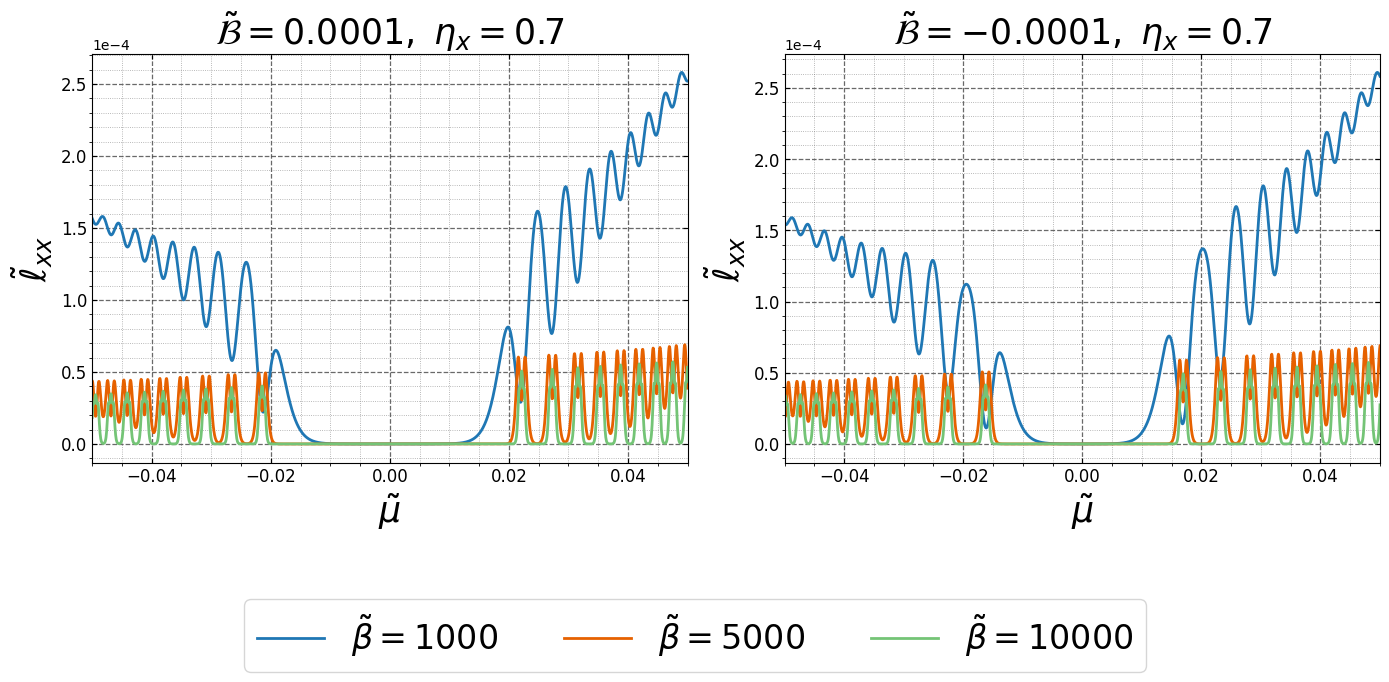}
\caption{Linear-response coefficients $\tilde \sigma_{xx}$, $\tilde \alpha_{xx}$, and $\tilde \ell_{xx}$, as functions of $\tilde \mu$. We have plotted them for 3 different values of $\beta $. We have taken $v_y/ v_x = 0.8 $.\label{fig:transport}}
\end{figure*}

\begin{figure*}[t!]
\centering
\includegraphics[width= 0.7 \textwidth]{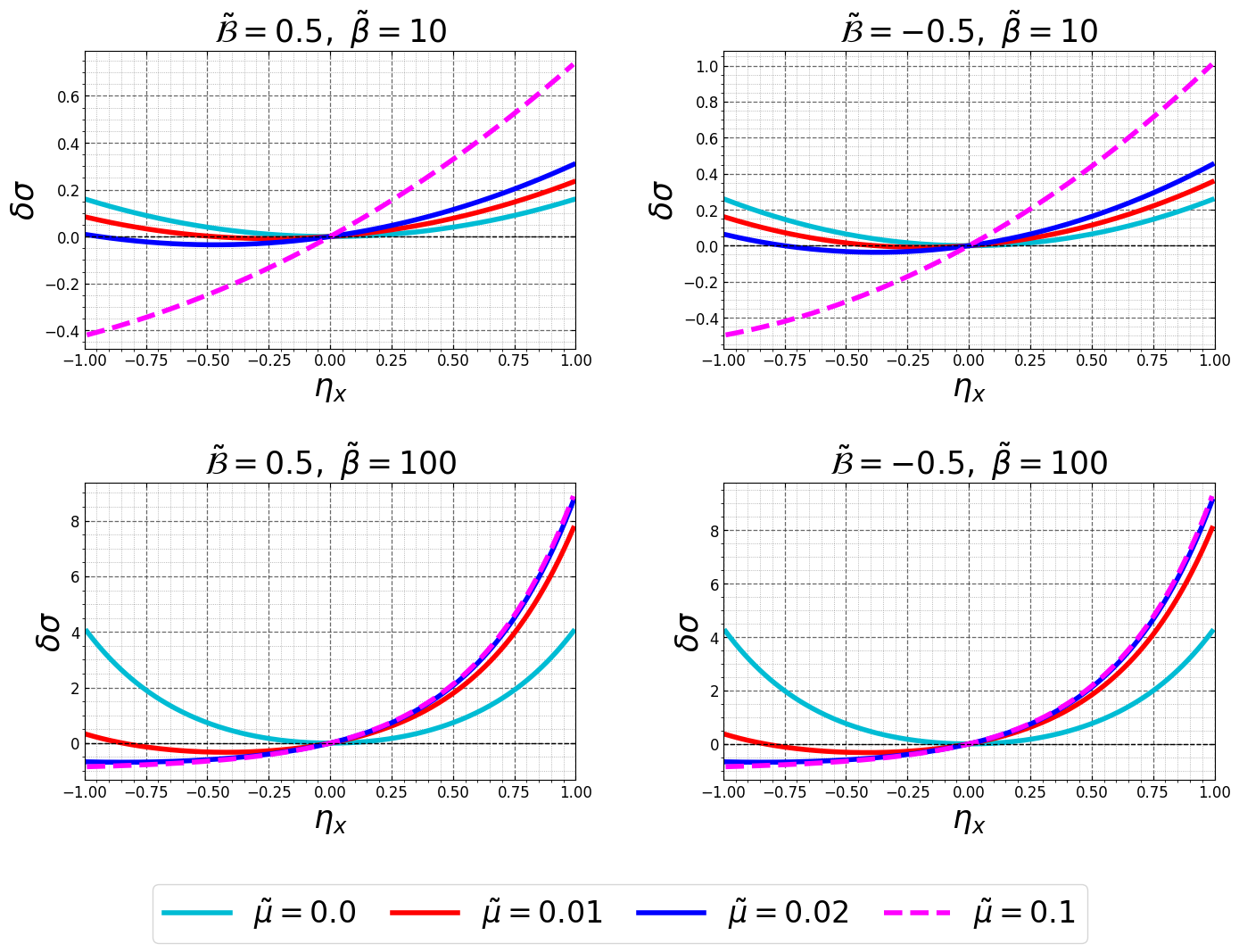}
\caption{Variation of electrical conductivity with $\eta_x$ by setting $v_y/ v_x$ to $ 0.8 $. Here, $ \delta \sigma \equiv \sigma_{xx}  (\eta_x)/ \sigma_{xx} (\eta_x=0) -1 $.\label{figsigeta}}
\end{figure*}

To find the solution for the PLLs of the full Hamiltonian in Eq.~\eqref{eqhtot}, we need to solve for the eigenvalue equation,
\begin{align}
\mathcal{H}_{\xi} \,\Phi (k_x, y) = E \,\Phi (k_x, y)\,,\quad
\Phi^T (k_x, y) = \begin{bmatrix}
\phi_A (k_x, y) & \phi_B(k_x, y) 
\end{bmatrix} .
\end{align}
Since $k_x$ remains a good quantum number (because of unbroken translation symmetry along the $x$-direction), we work with the momentum-variable for the $x$-direction. The steps for solving the coupled differential equations involving the unknowns, $\phi_A$ and $\phi_B$, are detailed in the Appendix. Following those intermediate steps, we get the PLLs as
\begin{align}
& \frac{  E^{\xi}_{n, \rm{si}} (k_x, \mathcal B) } {v_x} =  
\eta_x \,   k_x
+ \,\sqrt{\tilde r_\xi^2 \, \left( |\zeta | \, n
+ \frac{e \, \mathcal B+|\zeta|}{2} \right)^2
+ \left (k_x \, \tilde r_\xi + s_{-\xi} \, s_\xi \right )
\left (2 \, |\zeta |\, n + e \, \mathcal B+|\zeta| \right )}  \,,\nn
& \text{ where } \text{si} = \pm \,, \quad
s_\xi = 1 + \frac{\xi \, k_x\, a_0 } {2}\,, \quad
r_\xi = \frac{3}{2} + \frac{\xi \, k_x\, a_0 } {4}\,,\quad
\tilde{r}_\xi = a_0 \, r_\xi  \,, \quad \zeta =  e \, {\mathcal B} \, \frac{v_y}{v_x}\,,
\nn & \text{and } n \in \left\{ 0, \,1, \, 2,\cdots, 
\left\lfloor\frac{8 \,(2  + 3 \, \xi \,a_0\, k_x)}   { a_0^2\,|\zeta| \,(6 +\xi \,a_0\,  k_x)^2} \right\rfloor \right\}.
\label{eqevals}
\end{align}
We have labelled them by the integer-index $n$, momentum $k_x$, and band-index $\pm$ at the valley $\xi$. 
In the above expression, the symbol $\left \lfloor  \Upsilon \right \rfloor$ denotes the greatest integer less than or equal to $\Upsilon$.
We get an upper-bound for the number of PLLs that depends on strain induced through the pseudo magnetic field in the system. For small $ |\mathcal{B}| $, the expression approximates to
\begin{align}
\frac{ E^{\xi}_{n, \rm{si}} (k_x, \mathcal B) } {v_x} \simeq    \eta_x \,   k_x 
+ \text{si}\, \sqrt{\left (k_x \, \tilde r_\xi + s_{-\xi} \, s_\xi \right )
\left (2 \, |\zeta |\, n + e \, \mathcal B+|\zeta| \right )} .
\end{align}
Since $n\in \lbrace 0 \rbrace \cup \mathbb{Z}^+$, with an upper-bound of $
\left\lfloor\frac{8 \,(2  + 3 \, \xi \,a_0\, k_x)}   { a_0^2\,|\zeta| \,(6 +\xi \,a_0\,  k_x)^2} \right\rfloor$, we have a restricted number of PLLs. For positive values of $\eta_x$, it is possible for $E^\xi_{n, -} (k_x, \mathcal B)$
to take negative values as well. This happens when $ \eta_x \, a_0 \,  k_x  > \sqrt{\left (k_x \, \tilde r_\xi + s_{-\xi} \, s_\xi \right )
\left (2 \, |\zeta |\, n + e \, \mathcal B + |\zeta| \right )} $. Similarly, for $\eta_x<0$, we end up with negative values for $ E^\xi_{n, +}
( k_x, \mathcal B )$ as well.

\begin{figure*}[t!]
\centering
\includegraphics[width= 0.7 \textwidth]{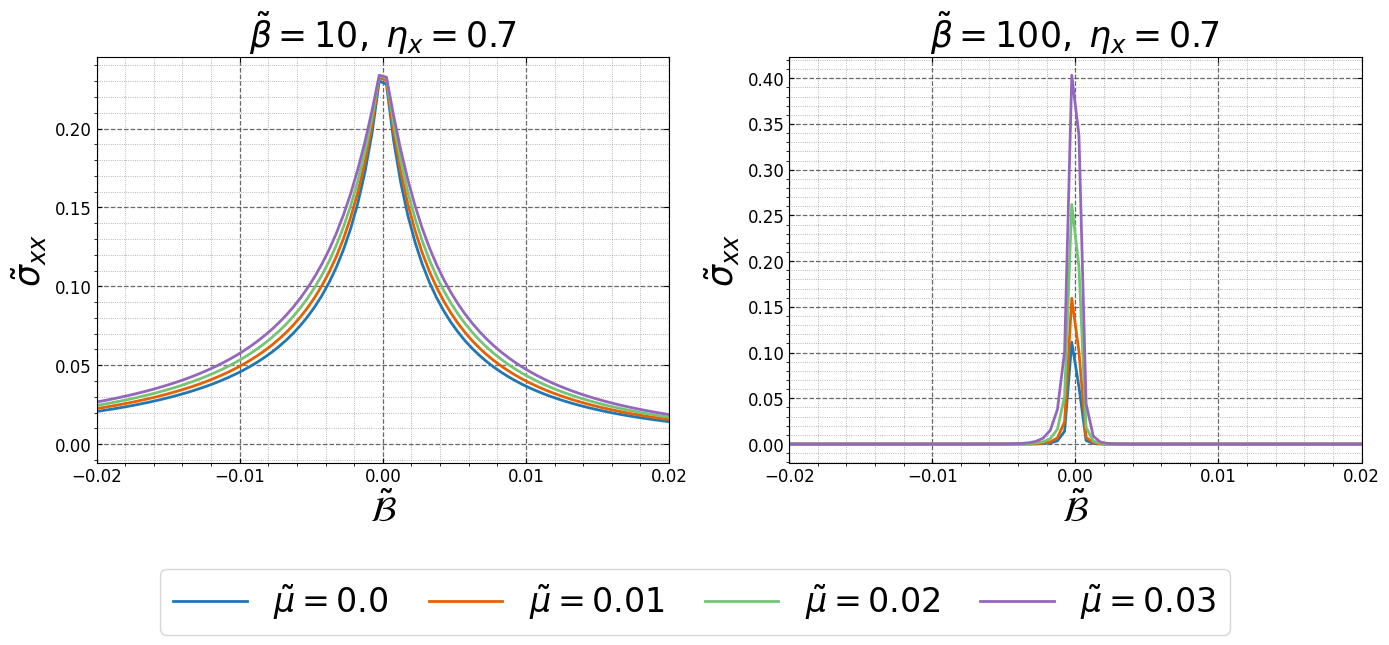}
\caption{Variation of electrical conductivity with $\mathcal B$.\label{figsigB}}
\end{figure*}

We represent the unique features of the PLLs by Fig.~\ref{fig:spectrum_B} for some representative parameter values. There, we observe the dispersive Landau levels, where the $\eta_x$-dependent term causes a tilt in the overall spectrum.
The most intriguing nature that is revealed comprises the PLLs forming closed curves rather than dispersing out indefinitely as $ |k_x|$ is cranked up. Of course there also exists a lone PLL in the form of a line (not closed) --- this one represents a non-dispersive or flat band in the absence of tilting [cf. the red lines in Fig.\ref{fig:spectrum_B}(b)]. Furthermore, the range of PLL-curves are more expanded for the $D$-valley compared to the $D^\prime$-valley, showing that the system is not valley-symmetric --- this is a consequence of the shifting of the Dirac points under strain.

\section{Transport Properties within the semiclassical Boltzmann formalism}

Having established the pseudo-Landau level (PLL) spectrum for tilted Dirac cones under strain, we now investigate the resulting transport properties within the semiclassical Boltzmann formalism. Unlike conventional Landau levels, the PLLs obtained here are apart from being dispersive are tilted as well in momentum $k_x$, which leads to qualitatively different transport behavior in set-ups shown in Fig.~\ref{fig:transport}.
In particular, the finite dispersion of the energy spectrum ensures a non-vanishing group-velocity along the $x$-direction, in contrast to the flat Landau levels of conventional quantum Hall systems. As a result, the system exhibits finite longitudinal conductivity in the bulk. The tilting of the PLL spectra provides an extra knob to tune the qualitative nature of the bulk linear response.


Within the semiclassical Boltzmann formalism, the longitudinal dc electrical conductivity is given by~\cite{arovas, ips-kush-review}
\begin{align}
\sigma^\xi_{xx} =  - \, \frac{e^2\,\tau} {L_y} \sum_{n, \, \text{si}} \int \frac{d k_x} {2 \,\pi}\,
\partial_{\varepsilon^{\xi, {\rm si }}_n} f_0 (\varepsilon^{\xi, {\rm si }}_n)
\, \left (\partial_{k_x }\varepsilon^{\xi, {\rm si }}_n \right)^2 ,
\end{align}
where $\tau$ is the relaxation time (taken to be momentum-independent in our simplified analysis), $ f_0  (\varepsilon) \equiv  \left[ e^{(\varepsilon - \mu)\,\beta} + 1\right ]^{-1} $ is the Fermi-Dirac distribution function, $ \varepsilon^{\xi, {\rm si }}_n \equiv \, 
 E^{\xi}_{n, \rm{si}} (k_x, \mathcal B)$, $\mu$ is the chemical potential, $\beta = 1/T $ is the inverse temperature, and $L_y$ is the length of the system in the $y$-direction. We take $\tau$ to be a phenomenological parameter describing various contributions to electronic scattering. The nonzero longitudinal conductivity is in stark contrast with the case of flat Landau levels, which yield vanishing longitudinal transport. 
We also compute the longitudinal thermoelectric conductivity and the longitudinal magnetothermal coefficient~\cite{arovas, ips-kush-review}, defined as
\begin{align}
\alpha^\xi_{xx} & =  \frac{e\,\tau} {L_y} \sum_{n,\,\text{si}} \int \frac{d k_x} {2 \,\pi}\,
\partial_{\varepsilon_n}  f_0 (\varepsilon^{\xi, {\rm si }}_n)\left (\partial_{k_x }\varepsilon^{\xi, {\rm si }}_n \right)^2  \frac{\varepsilon^{\xi, {\rm si }}_n - \mu}{T}
\text{ and } 
\ell^\xi_{xx} =  -\,\frac{\tau} {L_y} \sum_{n,\,\text{si}} \int \frac{d k_x} {2 \,\pi}\,
\partial_{\varepsilon^{\xi, {\rm si }}_n}  f_0 (\varepsilon^{\xi, {\rm si }}_n)\left (\partial_{k_x }\varepsilon^{\xi, {\rm si }}_n \right)^2  
\frac{ (\varepsilon^{\xi, {\rm si }}_n - \mu)^2}{T} ,
\label{eqalphaell}
\end{align}
respectively.

The explicit form of the group-velocity along the $x$-direction is captured by 
\begin{align}
v^{\xi, \text{si}}_x (n; k_x) \equiv \partial_{k_x}  E^{\xi}_{n, \rm{si}} (k_x, \mathcal B)  
=   \eta_x +\, \mathrm{si} \frac{\left( \mathcal B \, + (1 + 2n)\,|\zeta| \right)
\left(
\tilde{r}_{\xi} + k_x\, \tilde{r}_{\xi}
+ 2\, s_{\xi}\, s_{-\xi}
\right)
}  {2 \sqrt{
\left( \mathcal B \, + (1 + 2n)\,|\zeta| \right)
\left( k_x\, \tilde{r}_{\xi}\, + s_{-\xi}\, s_{\xi} \right)}}.
\end{align}
We note that the tilt-induced contribution enters the group-velocity as an additive term proportional to $\eta_x$, leading to a systematic shift ($\propto \eta_x^2 $) in the transport coefficients. This provides a direct way to tune response via strain-induced modifications of the bandstructure. 
For the ease of computation and plotting, we use the following variables: $ \tilde{\mathcal B }  = a_0^2 \, \mathcal B$, $\tilde \mu  = a_0 \, \mu/ v_x $, $\tilde k_x  = a_0 \, k_x $, $\tilde k_y =  a_0 \, k_y $, $\tilde y = y/a_0 $, $\tilde  E^{\xi}_{n, \rm{si}} (\tilde k_x, \tilde{\mathcal B} ) = a_0 \,  E^{\xi}_{n, \rm{si}} (k_x, \mathcal B) / v_x $, $\tilde \beta =  v_x\, \beta/a_0 $, $\tilde \sigma^\xi_{xx}  = L_y \, \sigma_{xx}^\xi / (\tau \, v_x ) $, $ \tilde \alpha^\xi_{xx}  = L_y \, \alpha_{xx}^\xi / (\tau \, v_x ) $, and $ \tilde \ell^\xi_{xx}  = a_0 \, L_y \, \ell_{xx}^\xi / ( \tau \, v_x^2 ) $. In what follows, we will focus on the valley with $\xi = + $ and, hence, we will remove the $\xi$-index everywhere for uncluttering the notations.

\begin{figure*}[t!]
\centering
\includegraphics[width=0.425 \textwidth]{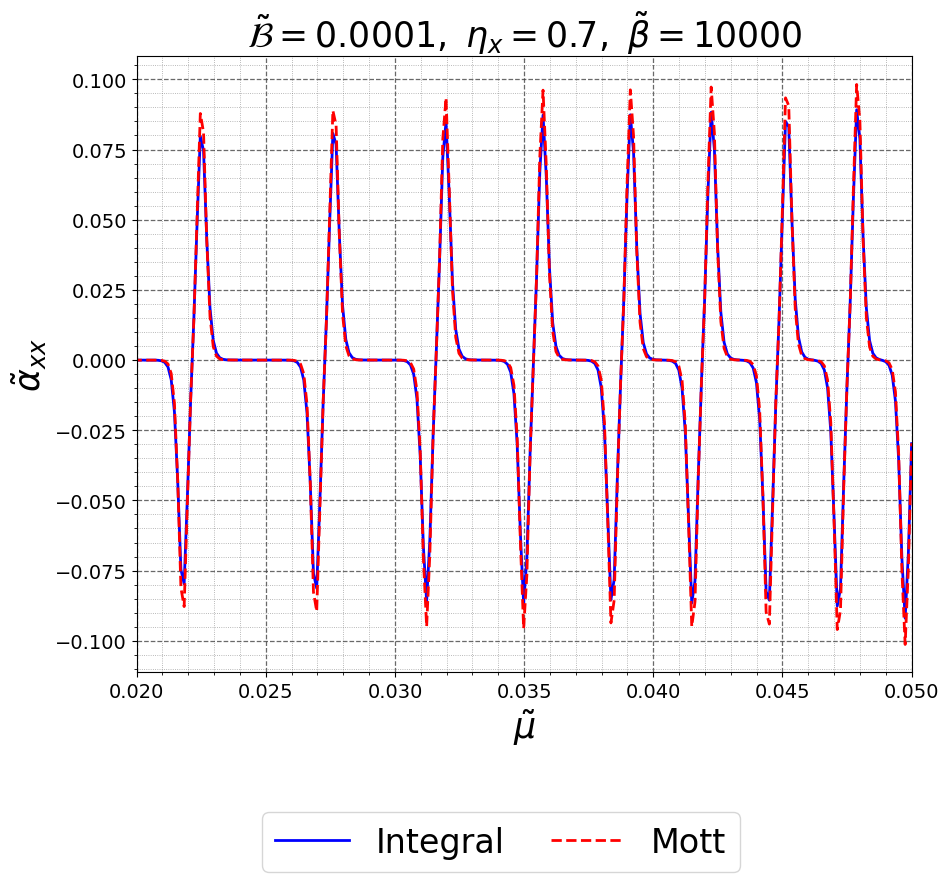}
\hspace{0.3 cm}
\includegraphics[width=0.42 \textwidth]{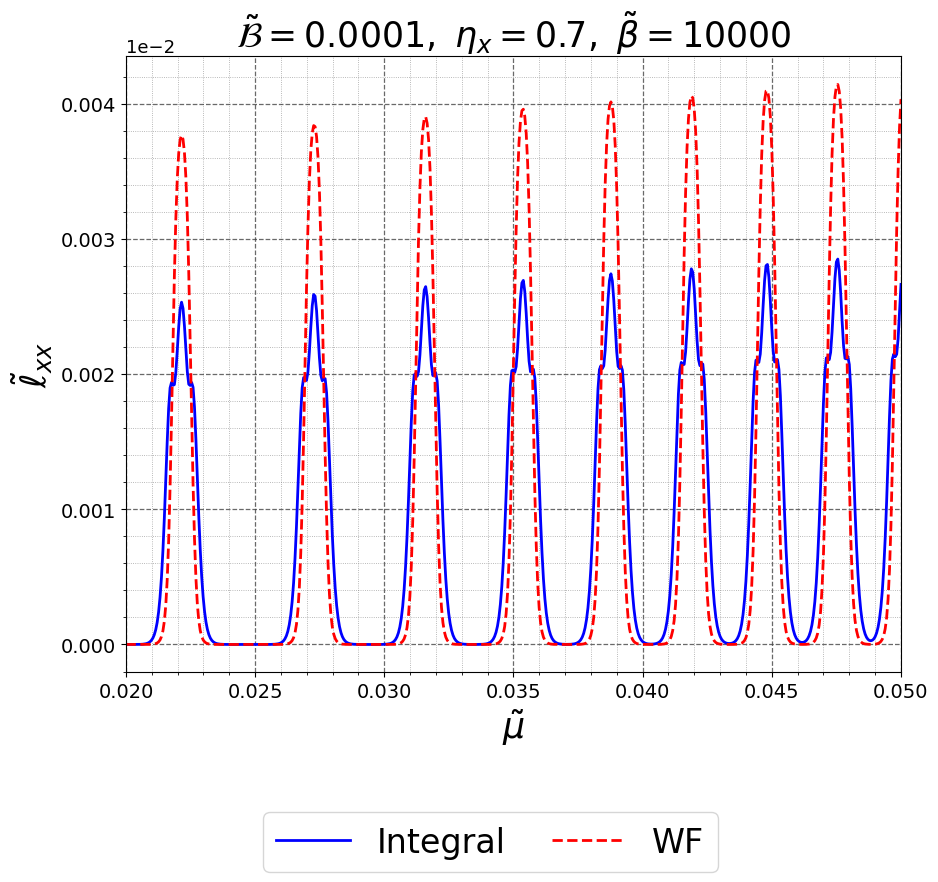} 
\caption{Comparision of the Mott Relation and the Wiedemann-Franz (WF) law [cf. Eq.~\eqref{eqmott}] with the results obtained from the original integrals [cf. Eq.~\eqref{eqalphaell}]. While the blue lines represent the results obtained from the actual integrals (indicated by the label ``Integral''), the dotted red lines show the answers obtained by using the relations. We have used here $v_y/v_x = 0.8$.\label{fig:compare}}
\end{figure*}

Some representative plots, as functions of $\mu$, are shown in Fig.~\ref{fig:transport} for several values of $\beta$. The electrical conductivity demonstrates sharp peaks as a function of $\mu$ --- albeit the peaks get progressively broadened as we go to higher temperatures (i.e., lower values of $\beta$) as the quantised bands of PLLs are exact only at $T = 0$ and a finite value of $T$ smears them out. Due to the extra factor of $(\varepsilon^{{\rm si }}_n - \mu)$ in the integral for $\alpha_{xx}$, compared to that for $ \sigma_{xx}$, each peak-like feature of the $\sigma_{xx}$-curves splits into a pair of asymmetric peaks in the $\alpha_{xx}$-curves. Analogously, due to the extra positive-definite factor of $(\varepsilon^{ {\rm si }}_n - \mu)^2 $ in the integral for $\ell_{xx}$, compared to that for $ \sigma_{xx}$, each peak-like feature of the $\sigma_{xx}$-curves splits into a pair of symmetric peaks in the $ \ell_{xx}$-curves. The response coefficients are asymmetric with respect to $\mathcal B = 0$, which is expected from the explicit forms of the PLLs [cf. Eq.~\eqref{eqevals}].
All these features are distinctly visible for higher and higher $\beta$-values, when thermal smearing gets minimised. 
The sensitivity on the values of $\eta_x$ has been depicted via Fig.~\ref{figsigeta}, where $ \delta \sigma \equiv \sigma_{xx}  (\eta_x)/ \sigma_{xx} (\eta_x=0) -1 $ has been plotted. The shifted-parabola shaped curves there show that the dependence has to be fitted with terms $\propto \eta_x$ and those $\propto \eta_x^2 $. At $\mu = 0$, we have unshifted parabolas (i.e., no $\propto \eta_x $ term). A nonzero $\mu$ always gvies a parabola with its vertex shifted away from the origin (towards negative values of both the horizontal and vertical axes) -- the shift increasing with increasing $|\mu|$. We also observe that, lower the temperature, higher the rate of change of the magnitude of $ \delta  \sigma  $ with an increasing $ |\eta_x| $-value.
Furthermore, we also illustrate the dependence of the strength of $ \sigma_{xx}$ on $\mathcal B $ through Fig.~\ref{figsigB}. We find that the response falls off rapidly with increasing values of $|\mathcal B |$, which can be traced back to the following: for the realistic values of $\mu $ we have used, the number of PLLs cut by the chemical potential, within a reasonable range of $k_x$ (which is the integration variable), decreases with rising $|\mathcal B|$.

Finally, we examine the validity of the Mott relation and the Wiedemann–Franz (WF) law in our strained system, which hold in metallic (or semimetallic) systems for $T  \rightarrow  0 $ in myriad situations \cite{ips-ruiz, ips-rsw-ph, ips-mwsm-floquet}. 
These two relations are captured by
\begin{align}
\label{eqmott}
\partial_{\mu} \sigma_{xx}\,(\mu)
= - \frac{3 \, e^{2}}{T \, \pi^{2}}\, \alpha_{xx} \, (\mu) \,
+ \mathcal{O}\,\left(T^2\right) \text{ and }
\sigma_{xx}\,(\mu)
= \frac{3 \, e^{2}}{T \, \pi^{2}}\, \ell_{xx}\,(\mu)\,
+ \mathcal{O}\,\left(T^2\right),
\end{align}
respectively.
In conventional metals, these relations hold under the assumption of a smooth energy-dependence of the conductivity near the Fermi level. 
Hence, \textit{a priori}, they are not guaranteed to hold for our case. To check their validity here, we compare the nature of the response-curves obtained by direct computation [via Eq.~\eqref{eqalphaell}] and those obtained from Eq.~\eqref{eqmott}. Such a numerical comparison is demonstrated in Fig.~\ref{fig:compare}, which shows that the relations produce a pretty good fit.

\section{Summary and discussion}
\label{secsum}

Contemporary advances in strain-engineering of two-dimensional materials have demonstrated remarkable control over electronic properties via lattice distortions. In particular, theoretical studies suggest that tailored strain in 2d as well as 3d materials can generate pseudomagnetic fields, giving rise to PLLs in the bulk and dispersive edge modes (that reflect the interplay between strain, confinement, and boundary effects). When applied to systems hosting Dirac-like dispersion in their pristine (i.e., unstrained) state, pseudomagnetic fields give rise to valley-dependent transport. The PLLs allow a significant departure from traditional quantum Hall physics, because longitudinal response arises due to the dispersive nature of the bands. The most intriguing feature of the pseudomagnetic field is that we are able to obtain an intrinsic deformation of the bandstructures without actually applying an external magnetic field.

We have focused on the formation of PLLs in tilted Dirac cones, extending the framework of Ref.~\cite{strained-dirac-franz} to incorporate tilt effects and, also, to determine the nature of the electric, thermoelectric, and thermal coefficients. We have calculated the longitudinal components of all the three kinds of response within the semiclassical Boltzmann formalism and have demonstrated their characteristics through representative curves (as functions of the relevant parameters). Our analysis reveals the modifications to expect in linear response, arising from the tilt-parameter. Indeed, if $\eta_x$ can be tuned continuously, it may lead to a mechanically configurable conductance through tilt-controlled residual velocity. Overall, our investigations show how anisotropy (including tilt) modify the  dispersion and transport properties of 2d Dirac-like systems. Our findings contribute significantly to the understanding of strain-engineering in 2d materials and open new pathways for controlling electronic properties through mechanical manipulations, which have potential implications for valleytronic devices and strain-tunable quantum technologies.

\appendix

\begin{widetext}
\section*{Appendix: Solving the differential equations}

In this section, we review the derivation of the eigenvalues and eigenvectors of the strained system [defined in Eq.~\eqref{eqhtot}], with a tilt with respect to the $k_x$-axis.
Starting with the Hamiltonian, 
\begin{align}
\label{eqhtotapp}
& \mathcal H_\xi (k_x, y) = \left[ 
 v_x \, \eta_x \, k_x \,  {\mathbb I}_{2\times 2}
 + \mathcal{H}_s(y; k_x, k_y) \right ]\big \vert_{k_y \rightarrow -i\, \partial_y}
\,, \quad 
\mathcal{H}_s  =   d^\xi_x (\bs k)\,\sigma_x +  d^\xi_y (\bs k)\,\sigma_y  \,, \nn &
\frac{ d^\xi_x(\boldsymbol{k}) } {v_x} = \xi \, k_x + s_{-\xi} \, e \, {\mathcal B} \, y\,,
\quad
\frac {d^\xi_y(\boldsymbol{k})} {v_y} =  k_y  \left(s_\xi - r_\xi \, e \, {\mathcal B} \, y\, a_0
\right )  ,
\quad s_\xi = 1 + \frac{\xi \, k_x\, a_0 } {2}\,, \quad
r_\xi = \frac{3}{2} + \frac{\xi \, k_x\, a_0 } {4} \,,
\end{align}
we follow the procedure detailed in Refs.~\cite{strained-dirac-franz, complex-LL-strained-graphene}. We solve the eigenvalue equation, $\mathcal H_{\xi=1} (k_x, y) = E\, \Phi (y; k_x)$, by choosing the Dirac-point $D$.

While dealing with the operators written in the $y$-space, we encounter terms of the form $i\, y\, \partial_y $, which are clearly not Hermiticised. One of the recipes of going from a classical operator (here, $ -i\, y\, \partial_y$) to its quantum version is symmetrisation, which we apply here (following Ref.~\cite{strained-dirac-franz}). Let us denote $y$ by $\check y$ and $-i\, \partial_y$ by $\check k_y$ to explicitly indicate that these are the position and momentum operators along the $y$-direction, projected into the position-space representation. Then we must replace any occurrence of $ -i\, y \,\partial_y $ by
$$ \frac{\check{y} \,\check{k}_y}{2}  +   \frac{\ \check{k}_y \, \check{y} }{2} = \check{y} \,\check{k}_y + \frac{1}{2} \,.$$
Going back to our original notation, we just replace $ -i\, y \,\partial_y $ by $-i\left ( y\,\partial_y + 1/2 \right)$ in the final expressions. After eliminating $\phi_A$ using the two equations arising from the two spinor components and shifting the $y$-coordinate as $ y\rightarrow y + 
\frac{s_+}{r_+ \,a_0\,  e\, {\mathcal B}}$, the final equation for $\phi_B$ turns out to be
\begin{align}
\label{eqdiffea}
& \left[(s_{-}\,e \, {\mathcal B}\,y)^2 + \frac{s_{-} \,e \, {\mathcal B}}{\tilde{r}}(2\,k_x\,\tilde{r}+2\,s_{-}\,s_+
-e \, {\mathcal B}\,\tilde{r}^2)\,y+\frac{\Delta}{\tilde{r}^2}-\frac{\zeta^2\,\tilde{r}^2}{4} \right] \, \phi_B 
-2 \,y \, \zeta^2 \, \tilde{r}^2 \, \partial_y \phi_B-y^2 \, \zeta^2 \, \tilde{r}^2 \, \partial^2_y \phi_B = 0\,,
\nn & \tilde{r} = a_0 \, r_+ \,, \quad \zeta =  e \, {\mathcal B} \, \frac{v_y}{v_x} \,, \quad
\epsilon = a_0 \, \left( \frac{E}{v_x}- \eta_x \, k_x \right), \quad
\Delta = (k_x \, \tilde r+ s_{-} \, s_+ )^2-\tilde r^2 \, \epsilon^2 \,.
\end{align}
The above form is in the form of a Sturm-Liuouville (SSL) problem, viz. $\partial_y \left[ \mathcal P (y)\,\partial_y \phi_B \right ] - \, {\mathcal Q} (y)\,\phi_B = -\,\lambda\, {\mathcal W}(y)\,\phi_B $. In fact, it is a singular SSL scenario because the interval $(-\infty, 0]$ (or, $ [0, \infty)$) infinite and ${\mathcal P} (y) = 0$ at the boundary $y= 0$. Next, we perform an asymptotic analysis of the system at $y\rightarrow0$ and $ |y| \rightarrow \infty$ Ref.~\cite{strained-dirac-franz}:
\begin{itemize}
\item For $y\rightarrow0$, the leading-order terms reduce to
\begin{align}
\left[\frac{\Delta}{\tilde{r}^2}-\frac{\zeta^2\,\tilde{r}^2}{4} \right] \, \phi_B 
-2 \,y \, \zeta^2 \, \tilde{r}^2  \, \partial_y \phi_B- \zeta^2 \, \tilde{r}^2 \,y^2 \, \partial^2_y \phi_B=0 \,,
\end{align}
which has solutions going as $\phi_B \sim y^{-\frac{1}{2}\pm{\frac{\sqrt{\Delta}}{\tilde{r}^2 \, |\zeta|}}} $. The $y=0$ regime represents one with regular singularity. The only physically-admissible solution is  $ \phi_B \sim y^{-\frac{1}{2}+{\frac{\sqrt{\Delta}}{\tilde{r}^2 \, |\zeta|}}} $, since the other one diverges. 

\item For $ |y| \rightarrow\infty$, we need to solve for
\begin{align}
\left (s_{-}\,e \, {\mathcal B}\,y \right )^2  \phi_B -y^2 \,\zeta^2 \, \tilde{r}^2 \, \partial^2_y \phi_B=0 \, ,
\end{align}
which has the solution of the form, $ y \sim \exp \left [ {\pm \, e\, |{\mathcal B}|  \,s_- \; y} /(\tilde r \, \zeta) \right ] $. 
Each solution diverges either at $ y = \infty$ or $y=-\infty $ and, hence, none is physically admissible along the entire $y$-axis. These points are therefore irregular singularities.

\end{itemize}
Knowing the forms of the asymptotic expressions, we now take the ansatz Ref.~\cite{strained-dirac-franz} (suppressing the explicit $k_x$-dependence),
\begin{align}
\phi_B (y) = e^{-\tilde y/2} \, 
y^{-\frac{1}{2}+{\frac{\sqrt{\Delta}}{\tilde{r}^2 \, |\zeta|}}} \,  u(y) \,,\quad
\tilde y = -\, \frac{ 2\, v_x \, \text{sgn} (\mathcal B) \,s_-} { v_y \,\tilde r }\, y \,.
\end{align}
Changing variables from $y$ to $\tilde y $ and defining $\tilde u(\tilde y) = u(y)$,
we end up with a confluent hypergeometric differential equation for $\tilde u(\tilde y)$, away from the singularity [considering the interval to be either $(-\infty, 0]$ or $ [0, \infty)$], which can be expressed as
\begin{align}
& {\tilde y}\, \partial^2_{\tilde y} \tilde u (\tilde y)
+ (\Gamma- \tilde y)\, \partial_{\tilde y} \tilde u({\tilde y})-\alpha \,\tilde u({\tilde y}) = 0\,, \quad
\alpha = \frac{\tilde r^2 \, (e \, {\mathcal B}+|\zeta|)+ 2 \, (\sqrt{\Delta}-s_{-} \, s_+ -k_x \, \tilde r)} 
{2 \, |\zeta| \, \tilde r^2} \,, \quad
\Gamma = 1+ \frac{2 \,\sqrt{\Delta }}{\tilde  r^2 \, |\zeta|}\, .
\end{align}
It turns out that the regular solutions involve the Kummer’s function, $\mathcal M (\alpha, \Gamma, \tilde y)$, where $\alpha = - n $ (taking $n \in 0 \cup {\mathbb{Z}}^+ $) is a non-positive integer but $\Gamma $ is not. Then, we have the condition that
\begin{align}
-\,n = \frac{\tilde r^2 \, (e \, {\mathcal B}+|\zeta|)+ 2 \, (\sqrt{\Delta}-s_{-} \, s_+ -k_x \, \tilde r)} 
{2 \, |\zeta| \, \tilde r^2} 
\Rightarrow
\sqrt{\Delta}=-\tilde r^2 \, |\zeta |\, n+k_x \, \tilde r-\frac{\tilde r^2}{2}
(e \, \mathcal B+ |\zeta|)+s_{-} \, s_+ \,.
\end{align}
Using the relation $ \Delta = (k_x \, \tilde r+s_{-} \, s_+ )^2-\tilde r^2 \, \epsilon^2 $, it leads to
\begin{align}
\epsilon^2 = \tilde r^2 \, \left( |\zeta |\, n + \frac{e \, \mathcal B+ |\zeta|}{2} \right)^2
+ \left (k_x \, \tilde r+s_{-} \, s_+ \right ) \left (2 \, |\zeta |\, n + e \, \mathcal B+|\zeta| \right),
\quad \mathcal M (-n, \Gamma, \tilde y) = {\mathcal L}^{\Gamma-1}_{n}(\tilde y)\,,
\end{align}
where ${\mathcal L}^\ell_{n} (\tilde y )$ is the Generalized (or, alternatively, Associated) Laguerre polynomial with $ n \in 0 \cup {\mathbb Z}^+$. Imposing the requisite physical condition that $\epsilon^2 \geq 0 $, we obtain an upper bound for $n$ as $ 
\left\lfloor\frac{8 \,(2  + 3 \, \xi \,a_0\, k_x)}   { a_0^2\,|\zeta| \,(6 +\xi \,a_0\,  k_x)^2} \right\rfloor  $. The symbol $\left \lfloor  \Upsilon \right \rfloor$ denotes the greatest integer less than or equal to $\Upsilon$. The final solution is obtained as Ref.~\cite{strained-dirac-franz}
\begin{align}
& \phi_B (k_x, y+ y^+_{\rm sg}) \, \Theta \left[ \text{sgn}(\mathcal B ) \left(- y\right)  \right],\,
\text{ where }\nn
 &\phi_B (k_x, y) = e^{\frac{  v_x \, \text{sgn} (\mathcal B) \,s_{-\xi}\, y} { v_y \, r_+\, a_0 }} \, 
y^{-\frac{1}{2}+{\frac{ v_x \,\sqrt{\Delta}}{ r_\xi^2\, a_0^2\, e  \, v_y  \left |{\mathcal B}  \right |}}} \, 
{\mathcal L}^{\Gamma-1}_{n}(\tilde y)\,, 
 \quad  y^+_{\rm sg} = \frac{s_+ } {r_+ \,a_0\,  e\, {\mathcal B}}\,,\quad
\tilde y = -\, \frac{ 2\, v_x \, \text{sgn} (\mathcal B) \,s_-} 
{ v_y \,r_+\, a_0 } \, y \,.
\end{align}
Clearly, the full wavefunction does not diverge at the irregular singularity, as it is cut-off by the Heaviside-theta function and it goes to zero before reaching that singular point.

The same process can be repeated for the valley with $\xi =- $. Using $E= \frac{v_x} {a_0} \left[  \eta_x \,   k_x + \epsilon   \right ]$, we finally get the PLLs as
\begin{align}
\frac{  E^{\xi}_{n, \rm{si}} (k_x, \mathcal B) } {v_x} =  
\eta_x \,   k_x + \text{si} \,
\sqrt{\tilde r_\xi^2 \, \left( |\zeta | \, n
+ \frac{e \, \mathcal B+|\zeta|}{2} \right)^2
+ \left (k_x \, \tilde r_\xi + s_{-\xi} \, s_\xi \right )
\left (2 \, |\zeta |\, n + e \, \mathcal B+|\zeta| \right )}  \,,
\text{ where } \tilde{r}_\xi = a_0 \, r_\xi \text{ and } \text{si} =\pm\,.
\end{align}
We have labelled them by the integer-index $n$, momentum $k_x$, and band-index $\pm$ at the valley $\xi$. The valley-dependent wavefunction, for the $n^{\rm th}$ PLL, is captured by
\begin{align}
& \phi_B^{\xi,n, \text{si}} (k_x, y+ y^\xi_{\rm sg}) \, \Theta \left[ \text{sgn}(\mathcal B ) \left(- y\right)  \right],\,
\quad
 \phi_B^{\xi, n, \text{si}} (k_x, y) = e^{\frac{ v_x \, \text{sgn} (\mathcal B) \,s_{- \xi}\, y} { v_y \, r_\xi\, a_0 }} \, 
y^{-\frac{1}{2}+{\frac{ v_x \,\sqrt{\Delta}}{ r_\xi^2\, a_0^2\, e  \, v_y  \left |{\mathcal B}  \right |}}} \, 
{\mathcal L}^{\Gamma-1}_{n}(\tilde y)\,, 
\quad  \Gamma = 1+ \frac{2 \,\sqrt{\Delta }}{\tilde  r_\xi^2 \, |\zeta|} \,,\nn
& \quad  y^\xi_{\rm sg} = \frac{s_\xi } {r_\xi \,a_0\,  e\, {\mathcal B}}\,,\quad
\tilde y = -\, \frac{ 2\, v_x \, \text{sgn} (\mathcal B) \,s_{-\xi}\, y} 
{ v_y \,r_\xi \, a_0 }\,, \quad
\sqrt{\Delta}=-{\tilde r}_\xi^2 \, |\zeta |\, n + k_x \, {\tilde r}_\xi -\frac{\tilde r_\xi^2}{2}
(e \, \mathcal B + |\zeta|)+s_{\xi} \, s_{-\xi} \,.
\end{align}
We also show below the explicit expression for the upper component of the $n^{\rm th}$-PLL's eigenspinor, $ \Phi^{\xi, n, \text{si}} (k_x, y) = \begin{bmatrix}
\phi_A^{\xi, n, \text{si}}  (k_x, y) && \phi_B^{\xi, n, \text{si}}  (k_x, y)  \end{bmatrix}^T $:
\begin{align}
\label{eqphia}
& \phi_A^{\xi, n, \text{si}} (k_x, y)  =  \frac{1}{v_x} \frac{
e^{\frac{2v_x \,\mathrm{sgn}(\mathcal{B})\, s_{-\xi}\, y}{v_y\, r_\xi\, a_0}}
\, y^{-\frac{3}{2} + 
\frac{v_x \sqrt{\Delta}}{r_{\xi}^2\, a_0^2\, e\, v_y\, \left|\mathcal{B}\right|}}
\left[
{\mathcal{F}}_1^\xi\,\mathcal{L}_{n}^{\Gamma-1}(\tilde{y})
-
{\mathcal{F}}_2^\xi \,\mathcal{L}_{n-1}^{\Gamma-1}(\tilde{y})\right]}
{ \frac{ E^{\xi}_{n, \rm{si}} (k_x, \mathcal B)} {v_x} - \eta_x \,   k_x }\,,
\nn & {\mathcal{F}}_1^\xi   = 8 \, v_x  \, F_1^\xi  \, F_2^\xi + 8 \,  r_\xi
\left [ 4 \,v_y|\mathcal{B}|\,r_\xi\,F_3^\xi 
+ 8  \, \mathcal{B} \, v_x \,  y\,F_4^\xi  \right ],\quad
{\mathcal{F}}_2^\xi  = 256\, v_x\, y\, |\mathcal{B}|\, r_\xi
\left[a_0 \,\mathcal{B} \,y \,r_\xi + v_y \left (s_\xi-2 \right )\right]
(s_\xi-2)\, \text{sgn}(\mathcal{B})\,,
\end{align}
where
\begin{align}
F_1^\xi  &=  2 \, \mathcal{B}  \, r_\xi^2 + a_0^2 \,  k_x^2 \left (1 - \xi \right )
-6 \, a_0  \, k_x - 4\,, \quad
F_2^\xi  = 4 \, a_0 \, \mathcal{B} \, y  \, r_\xi + 4 \, v_y\left (s_\xi-2 \right ), \nn
F_3^\xi  &= 4\,v_y \left (n+1 \right ) \left (s_\xi-2 \right ) 
+k_x  \,  \xi \,y \left[ a_0^3  \, \mathcal{B}\,(n+1)-4 \, v_x\right]
+6 \, a_0^2 \, \mathcal{B}\,(n+1)\,y\,, \nn
F_4^\xi  & = y\, r_\xi \left(4 \,a_0\,\mathcal{B}\,s_\xi-8\,a_0 \, \mathcal{B}
+ a_0 \,\xi\, v_y-2\, v_y \right)
+ 4\,v_y\left(s_{-\xi} \, s_\xi-1\right).
\end{align}

\end{widetext}


\bibliography{ref_ll}

\end{document}